
\magnification=1200
\overfullrule=0pt
\baselineskip=15pt
\font\mfst=cmr9
\font\mfs=cmr9 scaled \magstep 1
\font\rma=cmr9 scaled \magstep 2
\font\rmm=cmr9 scaled \magstep 3
\def\ueber#1#2{{\setbox0=\hbox{$#1$}%
  \setbox1=\hbox to\wd0{\hss$\scriptscriptstyle #2$\hss}%
  \offinterlineskip
  \vbox{\box1\kern0.4mm\box0}}{}}

\def\hn{$\;{\rm h}^{-1}$}

\def\R{\rm I\kern-.18em R}
\def\etal{{\it et al. }}
\topskip 6 true cm
\pageno=0

\centerline{\rmm OPTIMIZED LAGRANGIAN APPROXIMATIONS}
\smallskip
\centerline{\rmm FOR MODELLING LARGE--SCALE STRUCTURE}
\smallskip
\centerline{\rmm AT NON--LINEAR STAGES}
\bigskip\bigskip
\centerline{\rmm by}
\bigskip\bigskip
\centerline{\rmm T. Buchert$^{1}$, A.L. Melott$^{2}$, A.G.
Weiss$^{1}$}
\vskip 1.5 true cm
\centerline{\rma $^{1}$Max-Planck-Institut f{\"u}r Astrophysik}
\smallskip
\centerline{\rma Postfach 1523}
\smallskip
\centerline{\rma 85740 Garching, Munich}
\smallskip
\centerline{\rma F. R. G.}
\bigskip
\centerline{\rma $^{2}$Department of Physics and Astronomy}
\smallskip
\centerline{\rma University of Kansas}
\smallskip
\centerline{\rma Lawrence, Kansas 66045}
\smallskip
\centerline{\rma U. S. A.}

\vskip 1 true cm
\centerline{\mfs submitted to {\it Ap.J. Letters}}

\vfill\eject

\topskip= 4 true cm
\centerline{\rmm Optimized Lagrangian approximations}
\smallskip
\centerline{\rmm for modelling large--scale structure}
\smallskip
\centerline{\rmm at non--linear stages}
\bigskip\medskip
\centerline{\rma by}
\bigskip\medskip
\centerline{\rma T. Buchert, A.L. Melott, A.G. Weiss}
\vskip 0.8 true cm
\noindent
{\mfst
{\narrower
\baselineskip=12pt
{\mfs Summary:}
We report on a series of tests of Newtonian Lagrangian perturbation
schemes using N--body simulations for various power--spectra with
scale--independent indices in the range $-3$ to $+1$.
The models have been evolved deeply into the non--linear regime
of structure formation in order to probe the dynamical and
statistical performance
of the Lagrangian perturbation schemes (whose first--order solution
contains as a subset the celebrated ``Zel'dovich--approximation''
(hereafter
ZA). These tests reveal properties of the approximations at
stages beyond the obvious validity of perturbation
theory.
Recently, another series of tests of different analytical and
semi--numerical approximations for large--scale structure was
conducted with the
result
that ZA displays the best dynamical performance in comparison with the
N--body simulations, if the initial data were smoothed before
evolving
the model, i.e., a truncated form of ZA (TZA). We show in this Letter
that the
excellent performance of
TZA can be further improved by going to second order in the Lagrangian
perturbation approach.
The truncated second--order Lagrangian scheme provides
a useful improvement over TZA especially for negative power indices,
which suggests it will be very useful for
modelling standard scenarios such as ``Cold--'', ``Hot--'' and
``Mixed--Dark--Matter''.

}}

\vfill\eject
\topskip= 0 true cm
\baselineskip=14.5pt

\noindent{\rmm 1. Lagrangian perturbation theory put into perspective}
\bigskip\bigskip
\noindent
Zel'dovich (1970, 1973) proposed an approximation
(hereafter ZA) by extrapolating the Eulerian linear theory
of gravitational instability into the non--linear regime using
the Lagrangian picture of continuum mechanics. He discussed
interesting consequences of this approximation which is capable of
describing shell--crossing singularities which, in this model, develop
into highly anisotropic oblate ``pancake'' structures.
Those structures can be fundamentally understood and classified
in the framework of the Lagrange--singularity--theory,
(Arnol'd {\it et al.} 1982,
Shandarin and Zel'dovich 1989, Roshansky {\it et al.} 1994 and
references
 therein).
Zel'dovich's work initiated many
applications making it one of the most cited articles in
astronomy; it emerged as a standard tool to model principal elements
of the large--scale structure, is used to initialize most
N--body codes employed by
the cosmology community, and forms the basis of more sophisticated
approximations like the adhesion approximation (Gurbatov \etal 1989,
Kofman \etal 1992). Zel'dovich's model can be derived by formulating the
Euler--Poisson
system in terms of Lagrangian coordinates and solving the Lagrangian
evolution equations for the field of trajectories perturbatively
(Buchert and G\"otz 1987, Buchert
1989, 1992). It appears as a subclass of the irrotational Lagrangian
first--order solution which covers substantial non--linearities
in contrast to the Eulerian first--order solution. This
explains the success of this approximation if applied to non--linear
gravitational structure formation (compare Coles \etal 1993).

The particular justification that ZA could be relevant to hierarchical
clustering has developed slowly (Melott {\it et al.} 1983;
Melott and Shandarin
1990; Kofman 1991; Little {\it et al.} 1991;
Kofman {\it et al.} 1992; Coles
{\it et al.} 1993; however see Peebles 1993). A general concensus is
forming
based around a unification of the former Soviet (``pancake'')
and Western (hierarchical clustering) theories.

The Lagrangian theory of gravitational instability
is now used in large--scale structure modelling
much as the Eulerian theory of gravitational instability used to be.
Numerous efforts concern the investigation and application of Lagrangian
perturbation solutions up to the third order
(Buchert 1989, 1992, Moutarde \etal 1991,
Bouchet \etal 1992, Buchert 1993, Buchert and Ehlers 1993, Gramann
1993, Giavalisco \etal 1993, Lachi\`eze--Rey 1993a,b, Buchert 1994,
Juszkiewicz \etal 1994, Bernardeau 1994, Munshi and Starobinsky 1994,
Munshi \etal 1994), and, most
recently, the investigation of general relativistic analogues (which are
intrinsically Lagrangian in the eigensystem of the flow)
(Matarrese \etal 1993, Kasai 1993, Croudace \etal 1994,
Bertschinger and Jain 1994, Matarrese \etal 1994a,b, Berschinger and
Hamilton 1994, Kofman and Pogosyan 1994, Salopek \etal 1994).

\vfill\eject

\noindent{\rmm 2. Optimization of Lagrangian perturbation schemes}
\bigskip\medskip
\noindent
Until recently, ZA was evolved only until shell--crossing, i.e., when
singularities in the density field develop, at the epoch when the
Eulerian
representation of the basic dynamical equations
breaks down. In principle, the Lagrangian representation of
the flow allows following the evolution across caustics, where the
flow field itself remains finite. This implies
neglecting self--gravitating interaction of multi--stream systems
developing inside caustics, however, secondary generations of
shell--crosssings can be modelled as observed in N--body simulations by
going to higher orders in the perturbation approach (Buchert and Ehlers
1993). Melott \etal 1994a (hereafter MPS) investigated
the performance of a new approximation which requires truncation of high
frequencies in the initial power--spectrum before evolving ZA
(hereafter: TZA) taking the evolution of large--scale
structure deeply into the non--linear regime. MPS found that filtering
the
initial
data with
a Gaussian at a scale close to but smaller than the non--linearity scale
yields the best agreement with the density fields of the same
(untruncated) initial data as evolved by an
N--body code.

The non--linearity scale
$k_{nl}$ is defined by:
$$
a^2(t) \; \int_0^{k_{nl}(t)} \; d^3 k \; {\cal P}(k)
\;=\;1\;\;,\eqno(1)
$$
where $k_{nl}(t)$ is decreasing with time as successively larger scales
enter the non--linear regime; $a(t)$ is the scale factor of the
homogeneous background ($a(t_i)\equiv 1$), and ${\cal P}(k)$ denotes the
initial power--spectrum taken to be a powerlaw with indices in the range
$-3$ to $+1$.

``Best agreement'' was defined in terms
of an optimal scale $k_{opt}$ in $k$--space at which the
usual cross--correlation coefficient $S$ between the resulting density
fields attains its maximum:
$$
S := {<(\delta_1 \delta_2)> \over \sigma_1 \sigma_2} \;\;, \eqno(2)
$$
where $\delta_{\ell}, \ell=1,2$  represent the density contrasts in
the analytical and the numerical approximations, respectively,
$\sigma_{\ell} = \sqrt{<\delta_{\ell}^2>-<\delta_{\ell}>^2}$
is the standard deviation in a Gaussian random field;
averages $<...>$ are taken over the entire distribution. We believe this
is the
most important statistical test, because it measures whether the
approximation
is moving mass to the right place, with an emphasis on dense regions. We
also allow
for small errors by calculating $S$ for the two density arrays
smoothed at a
variety of smoothing lengths.

For the Lagrangian perturbation schemes up to the third order which were
used in our tests see (Buchert 1994). We conducted several
tests: In the first step we studied ``pancake models'', i.e., models
which a priori have a truncated power--spectrum, in order to study
principal effects of a second-- and higher--order correction
to ZA (for details see Buchert \etal 1994a).
In the second step we analyzed the whole family of models
with powerlaw--spectra $-3,\ldots,+1$ by evolving them deeply into the
non--linear regime (for details see Melott {\it et al.} 1994b).
These ``hierarchical models'' have been evolved for expansion
factors of 240 to 5100, depending on spectral index and $k_{nl}$.

Besides the cross--correlation coefficient (3) as a function of scale,
we analyzed several statistics including the comparison of the
evolved power--spectrum, the evolved r.m.s. values of the density
contrast as a function of scale, the phase--angle accuracy achieved by
the analytical models, and the evolved density distribution functions.

For all these statistics and for all spectra studied with different
filter types and filter scales, we always found improvement for the
second--order
scheme upon first--order (TZA), if the initial data are truncated
with a {\it Gaussian filter} at a slightly larger scale $k_{opt}$ than
the scale
needed for the optimal TZA.

In Fig. 1 we display slices of the final density fields for the
spectrum with index $n=-1$ as calculated
by the N--body code and the first-- and second--order Lagrangian
approximations at the stage where
the non--linearity scale has evolved to $k_{nl}=8k_f$; $k_f$ is
the fundamental mode of the simulation box.

In Fig. 2 we present the results of the
most important statistical test
which probes the dynamics of the models, i.e., the cross--correlation
coefficient (3)  for the whole family of hierarchical models
at the same evolution stage.

Considerably more details can be found in Melott {\it et al.} (1994b),
where
we show more statistical tests applied for the range of indices $-3\leq
n\leq
 1$.
\vfill\eject

\noindent{\rmm 3. Conclusions}
\bigskip\medskip
\noindent
We summarize our main conclusions and list the advantages of going to
second--order
in perturbation theory for the purpose of modelling highly non--linear
stages:
\bigskip
$\bullet$
1. The statistics which probe
the gravitational dynamics of the models show improvement due to
second--order corrections. This success is found for a considerably
higher non--linearity than expected from a perturbation approach.

$\bullet$
2. The improvement (although minor for much
small--scale power) is {\it robust} by going to later stages and
to smaller scales. This holds for any spectrum and for any
statistics analyzed.

$\bullet$
3. The CPU times on a CRAY YMP are for the first--order scheme $25$
seconds, and for the second--order scheme $60$ seconds;
the corresponding CPU times on a CONVEX C220 are $2$ and $5$ minutes.
Thus, even the second--order
scheme is competitive with {\it one step} in a corresponding
PM--type N--body simulation.

$\bullet$
4. The high speed as well as the fact that the second--order scheme is
as
easy to implement as the first--order scheme (directly from the initial
data), render this model suitable for all areas of application where
thus far ZA was used, e.g., the initialization of N--body codes.

$\bullet$
5. The second--order scheme predicts much faster collapse of first
objects (treating also tidal effects) at times comparable to the
collapse time in the
widely used spherical ``tophat'' model (Moutarde \etal 1991, Buchert and
Ehlers 1993, Munshi \etal 1994). Thus, it is preferred for the treatment
of ensembles of collapsing objects and for normalization purposes.

$\bullet$
6. Since the second--order corrections to TZA provide noticeable
improvement of dynamical accuracy for initial data
with negative sloped power--spectra, we expect that
the truncated second--order scheme will be especially useful for the
modelling
of standard cosmogonies (like Hot--, Cold--, and Mixed--Dark--Matter).

$\bullet$
7. This modelling will be effective for large sample
calculations, since in numerical realizations
of `fair' samples in excess of $300$\hn Mpc, performed with
the same resolution as the simulations reported here ($128^3$ particles
on $128^3$ meshs), the truncation scale is close to the Nyquist
frequency of the
N--body computing. Thus, shortcomings of the analytical schemes become
negligible which puts them in an ideal position for the purpose of
simulating the environment of galaxy formation down to scales
where other physical effects start to affect models based
on the description of self--gravity alone.
Our method can be effective down to
galaxy group mass scales ($10^{13} M_\odot$),
or better if we include biasing or
go to epochs earlier than the present. Thus many things which have been
studied
by N--body simulation can now be generated by approximation.
The code for second--order is available on request from
tob @ mpa-garching.mpg.de $\;$.

$\bullet$
8. The third--order scheme does not show the `robustness' observed for
second-- and first--order. However, to draw definite conclusions the
{\it analytical} solution of the third--order effect must be studied
with
 reduced
numerical uncertainties in its realization by Fast--Fourier--Transform,
as
pursued by Buchert \etal (1994b).

\smallskip\bigskip\noindent
{\rma Acknowledgements:} {\mfst
TB is supported by DFG (Deutsche Forschungsgemeinschaft). ALM wishes to
acknowledge support from NASA NAGW--3832 and from NSF grants
AST--9021414 and
OSR--9255223, and facilities of the National Center for Supercomputing
Applications, all in the USA.
We are grateful to the Aspen Center for Physics
(USA) for its June 1994 workshop on topics related to this work.}

\vfill\eject

\def\ref{\par\noindent\hangindent\parindent\hangafter1}
\centerline{\rmm References}
\bigskip\bigskip
{\mfst
\ref
Arnol'd, V.I., Shandarin, S.F., Zel'dovich, Ya.B. 1982 Geophys
Astrophys Fluid Dyn  20, 111
\smallskip
\ref
Bernardeau, F. 1994 ApJ, in press
\smallskip
\ref
Bertschinger, E., Jain, B. 1994 ApJ, in press
\smallskip
\ref
Bertschinger, E., Hamilton, A.J.S. 1994 ApJ, submitted
\smallskip
\ref
Bouchet, F.R., Juszkiewicz, R., Colombi, S., Pellat, R. 1992
ApJ Lett  394, L5
\smallskip
\ref
Buchert, T., G\"otz, G. 1987 J Math Phys 28, 2714
\smallskip
\ref
Buchert, T. 1989 Astron Astrophys  223, 9
\smallskip
\ref
Buchert, T. 1992 MNRAS  254, 729
\smallskip
\ref
Buchert, T. 1993 Astron Astrophys Lett 267, L51
\smallskip
\ref
Buchert, T., Ehlers, J. 1993 MNRAS  264, 375
\smallskip
\ref
Buchert, T. 1994 MNRAS, in press
\smallskip
\ref
Buchert, T., Melott, A.L., Weiss, A.G. 1994a Astron
Astrophys, in press
\smallskip
\ref
Buchert, T., Karakatsanis, G., Klaffl, R., Schiller, P. 1994b
in preparation
\smallskip
\ref
Coles, P., Melott, A.L., Shandarin, S.F. 1993 MNRAS
 260, 765
\smallskip
\ref
Croudace, K.M., Parry, J., Salopek, D.S., Stewart, J.M. 1994
ApJ, in press
\smallskip
\ref
Giavalisco, M., Mancinelli, B., Mancinelli, P.J., Yahil, A. 1993
ApJ  411, 9
\smallskip
\ref
Gramann, M. 1993 ApJ Lett  405, 47
\smallskip
\ref
Kasai, M. 1993 Phys Rev  D47, 3214
\smallskip
\ref
Kofman, L.A. 1991 in Primordial Nucleosynthesis and Evolution of the
Universe
(K. Sato and J. Audouze) Kluwer Scientific Dordrecht
\smallskip
\ref
Kofman, L.A., Pogosyan, D., Shandarin, S.F., Melott, A.L. 1992
ApJ  393, 437
\smallskip
\ref
Kofman, L.A., Pogosyan, D. 1994 ApJ, submitted
\smallskip
\ref
Lachi\`eze-Rey, M. 1993a ApJ  407, 1
\smallskip
\ref
Lachi\`eze-Rey, M. 1993b ApJ  408, 403
\smallskip
\ref
Little, B., Weinberg, D.H. and Park, C.B. 1991 MNRAS 253, 295
\smallskip
\ref
Matarrese, S., Pantano, O., Saez, D. 1993 Phys Rev
D47, 1311
\smallskip
\ref
Matarrese, S., Pantano, O., Saez, D. 1994a Phys Rev Lett 72, 320
\smallskip
\ref
Matarrese, S., Pantano, O., Saez, D. 1994b MNRAS, submitted
\smallskip
\ref
Melott, A.L., Einasto, J., Saar, E., Suisalu, I., Klypin, A.A. and
Shandarin,
S.F. 1983 Phys Rev Lett 51, 935
\smallskip
\ref
Melott, A.L. and Shandarin, S.F. 1990 Nature 346, 633
\smallskip
\ref
Melott, A.L. 1994 ApJ, in press
\smallskip
\ref
Melott, A.L., Pellman, T.F., Shandarin, S.F. 1994a MNRAS,
in press
\smallskip
\ref
Melott, A.L., Buchert, T., Weiss, A.G. 1994b Astron Astrophys,
submitted
\smallskip
\ref
Moutarde, F., Alimi, J-M., Bouchet, F.R., Pellat, R., Ramani, A.
1991 ApJ  382, 377
\smallskip
\ref
Munshi, D., Starobinsky, A.A. 1994 ApJ, in press
\smallskip
\ref
Munshi, D., Sahni, V., Starobinsky, A.A. 1994 ApJ, submitted
\smallskip
\ref
Peebles, P.J.E. 1993 Principles of Physical Cosmology
(Princeton: Princeton University Press) Chapter 22
\smallskip
\ref
Roshansky, L.W., Shandarin, S.F., Buchert, T., Bartelmann, M. 1994,
in preparation
\smallskip
\ref
Salopek, D.S., Stewart, J.M., Croudace, K.M. 1994 MNRAS,
submitted
\smallskip
\ref
Shandarin, S.F., Zel'dovich, Ya.B. 1989 Rev Mod Phys
61, 185
\smallskip
\ref
Zel'dovich, Ya.B. 1970 Astron Astrophys  5, 84
\smallskip
\ref
Zel'dovich, Ya.B. 1973 Astrophysics  6, 164

}

\bigskip\bigskip
\bigskip\bigskip

\centerline{\rmm Figure Captions}
\bigskip\medskip
{\mfst
\noindent{\bf Figure 1:} Thin slices
(thickness $L/128$) of the
density fields are displayed for the numerical (a), the
optimally truncated first--order
(b) and second--order (c) approximations for
the evolution stage corresponding to $k_{nl}=8 k_f$, and for
the single power--spectrum with index $n=-1$.
The grey--scale is logarithmic
in order to emphasize the high--density regions.

\bigskip\smallskip
\noindent{\bf Figure 2:} The cross--correlation coefficient $S$
as a function of the standard deviation $\sigma_{\rho}$ of the
smoothed numerical simulation for
the different power--spectra $n=-3,-2,-1,0,+1$ (Figs. 2a,b,c,d,e). The
cross--correlation of the N--body
with the optimally truncated first--order model is
shown as a dotted line; with the optimally truncated second--order
model a dashed line.

}
\vfill\eject
\bye